\documentstyle[prl,aps,epsf,cite]{revtex}

\newcommand{\newc}{\newcommand}
\newc{\beq}    {\begin{equation}}
\newc{\eeq}    {\end{equation}}
\newc{\beqa}    {\begin{eqnarray}}
\newc{\eeqa}    {\end{eqnarray}}
\newc{\no}    {\\ \nonumber}

\def\PLA{{ Phys. Lett.} { A} }

\def\PRL{{ Phys. Rev. Lett. }}

\def\PRA{{ Phys. Rev.} { A} }

\begin{document}


\twocolumn[\hsize\textwidth\columnwidth\hsize\csname
@twocolumnfalse\endcsname

\title{Quantum Cryptography Using Single Particle Entanglement}
\author{ Jae-Weon Lee and  Eok Kyun Lee  \\}
\address{
 Department of Chemistry,  School of Molecular Science (BK 21),
Korea Advanced   Institute of Science and Technology,  Taejon
 305-701, Korea.}

\author{  Yong Wook Chung and Hai-Woong Lee}
\address{
Department of Physics, Korea Advanced Institute of Science and
Technology, Taejon 305-701, Korea \\}

\author{ Jaewan Kim }
\address{
School of Computational Sciences,
Korea Institute for Advanced Study,
207-43 Cheongryangri-dong, Dongdaemun-gu Seoul 130-012, Korea}

\date{\today}
\maketitle

\begin{abstract}
{\normalsize
A quantum cryptography scheme based on entanglement between
a single particle state
and a vacuum state is proposed.
The scheme utilizes linear optics devices to detect
the superposition of the vacuum and single particle states.
Existence of an eavesdropper can be
detected by using a variant  of 
Bell's inequality.}
\end{abstract}
\pacs{PACS:03.65.BZ, 42.50.Dv,42.50.Ar}
\maketitle
]
Entanglement could  be exploited in 
many interesting applications, including
quantum teleportation\cite{PRL70_1895,N390_575} and
quantum cryptography\cite{PRL67_661}.
Discussion on the nonlocal nature (entanglement) of
quantum systems was initiated by Einstein, Podolsky and Rosen
(EPR)\cite{PR47_777} and later extended by
 Bell\cite{PLIC1_195,RPP41_1881,PRL49_91}. Since then
 many authors  have studied the physical
meaning
of the nonlocality of a single particle\cite
{bjork,Tan,hardy,santos,mann,gerry,michler,leekim}.
Generally, quantum cryptography schemes based on entanglement
 (EPR-based schemes) use two or more spatially separated particles
 possessing correlated properties  as the source
of entanglement. However, recent developments in 
 experimental techniques\cite{PRL89,PRA62,experiment}
 for generating and manipulating single photons
have made quantum information processing
utilizing single particle entanglement feasible.
Here, single particle entanglement refers to
entanglement of a single particle state and the vacuum state
\cite{czachor}.

In the present study, we developed 
a quantum cryptography  scheme 
based on 
single particle entanglement. The propopsed scheme utilizes
linear optics to detect a superposition
of the vacuum state and a single photon state.
A variant of Bell's inequality suggested by Peres\cite{peres}
is used for the detection of eavesdropping.
In fact, the idea of quantum cryptography using single particle
entanglement is not new.
Examples of other approaches that can be considered as
quantum cryptography schemes using single particle
entanglement are the
phase coding scheme of Bennett\cite{bennett}
and Ardehali's scheme based on the delayed choice experiment\cite{ardehali},
which uses interferometers.
In these double-rail schemes, detection of a particle state
is performed by a single observer at a given site.
 A characteristic feature of our
single rail scheme is that both of two space-like separated parties,
whom we call Alice and Bob, detect either a 
single particle or no particle at their respective sites.
This characteristic makes our scheme more compatible with the
original meaning of quantum nonlocality.

We begin with a description of
 our scheme, which is depicted in Fig. 1.
 The setup consists of a single photon source (S) and a lossless 50/50 beam
splitter ($BS_0$), which generate
the single particle entanglement state,
and two identical non-deterministic projective measurement devices belonging to
 Alice and Bob, respectively. 
Each projective measurement device shown in detail in Fig. 2
itself consists of a lossless 50/50 beam splitter
 ($BS_A$ or $BS_B$) with a probe state $\gamma|0\rangle
+\delta|1\rangle$  and two photon detectors ($D_{Aa},D_{Ab}$ or $D_{Ba},D_{Bb}$).
We assume that every beam splitter induces a sign change in a transmitted beam
incident on the black side (Eq. (\ref{trans})).

\begin{figure}[Fig1]
\epsfysize=6cm \epsfbox{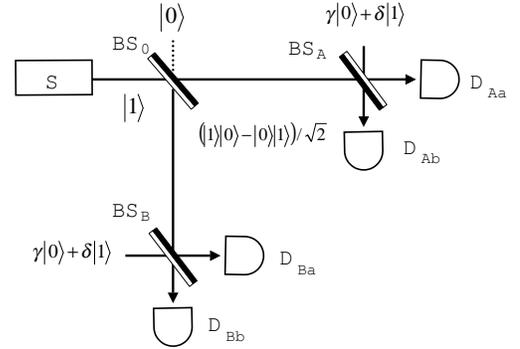}
 \caption[Fig1]{\label{fig1}
 Schematic of the experimental setup
  for quantum cryptography based on single particle
entanglement.
See text for a detailed explanation.
}
\end{figure}
 The output
state emerging from the beam splitter $BS_0$ is  given by
(see Eq. (\ref{psi}))
\begin{equation}
|\phi\rangle=\frac{1}{\sqrt{2}}\left (|1\rangle_A|0\rangle_B
-|0\rangle_A|1\rangle_B \right ) \label{inputstate},
\end{equation}

where subscripts A and B refer to the modes of the photons exiting the
beam splitter through the output ports A (towards Alice) and
B (towards Bob), respectively,
and $|1\rangle$ and $|0\rangle$ are the
single photon state and the vacuum state, respectively.

\begin{figure}[Fig2]
\epsfysize=6cm \epsfbox{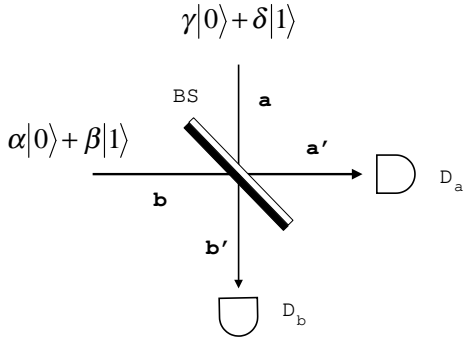}
 \caption[Fig2]{\label{fig2}
 Schematic of a device for performing a
  nondeterministic projective measurement
  of the superposition state of the vacuum and a single photon
  $\alpha|0\rangle+\beta|1\rangle$. $\gamma|0\rangle+\delta|1\rangle$
  is a known probe state.
}
\end{figure}
The state given in Eq. (\ref{inputstate}) represents a single-photon
entangled state.
Following the argument of Peres\cite{peres} , Alice and Bob, who test a
  violation of the Bell's inequality,
measure the projection  on the superposed state of
a single particle and the vacuum $\alpha|0\rangle+\beta|1\rangle$.
However, detection of the superposition 
of a particle state and the vacuum state is made difficult by the fact that the superposed
state is not a particle number eigenstate. 
 The experimental setup shown in Fig. 2, which is a generalization of
 the setup  considered in Ref. \citen{lund},
 can be used to detect the superposed state.
The beam splitter $BS$ (corresponding to the beam splitter
 $BS_A$  or $BS_B$ in Fig. 1)
performs the mode transformation
\beq
\label{trans}
  \left (
         \begin{array}{cc}
          a'\\
          b'
           \end{array}
    \right )=\left (
         \begin{array}{cc}
          \sqrt{R} & \sqrt{1-R}  \\
          -\sqrt{1-R} & \sqrt{R}
           \end{array}
    \right )
    \left (
         \begin{array}{cc}
          a\\
          b
           \end{array}
    \right ),
\eeq
where $R$ is the reflectivity of the beam splitter.
Using  second quantized notation, the general form of the input state
shown in  Fig. 2 can be
written  as
\beq
\label{bs1}
\psi=(\gamma+\delta a^{\dagger})
(\alpha+\beta b^{\dagger})|0\rangle
\eeq
with normalization requirements $\gamma^2+\delta^2=1$ and $\alpha^2+\beta^2=1$.
Here, $\gamma|0\rangle+\delta|1\rangle$ is a known probe state
with fixed $\gamma$ and $\delta$, while
$\alpha|0\rangle+\beta|1\rangle$ is an unknown state to be
measured. The probe state can be prepared by linear optics with coherent
light and a single photon state\cite{lund} or
by parametric down conversions\cite{hardy}.
By replacing $a$ and $b$ in
Eq. (\ref{bs1}) with $a'$ and $b'$ obtained from the transformation
Eq. (\ref{trans}),
 we obtain the following output state
\beqa
\label{psi}
\psi&=&[
  \alpha \gamma  + {\sqrt{R(1 - R)}}\beta
   \delta (a'^{\dagger2}- b'^{\dagger2})
   -\beta \delta( 1-2R  ) a'^\dagger b'^\dagger \no
   &+&
  ( \sqrt{1 - R}\beta \gamma+
     \sqrt{R}\alpha \delta ) a'^{\dagger} \no
     &+&( {\sqrt{R}}\,\beta \,\gamma  -
     {\sqrt{1 - R}}\,\alpha \,\delta ) b'^{\dagger}
     ]|0\rangle
\eeqa
Hence, by setting  $R=1/2$ and choosing
$\gamma$ and $\delta$ which satisfy
\beq
\label{condition}
\alpha\delta= \beta \gamma,
\eeq
one finds that the coefficient of the $b'^{\dagger}$ term vanishes
while that of the $a'^{\dagger}$ term does not.
In other words, there is a possibility that
detector $D_a$ detects a single photon, while $D_b$ detects none.
By noting this event,
one can perform a nondeterministic projection on the
 superposition state $\alpha|0\rangle
+\beta|1\rangle$.
Using the parameters chosen above, the output state can be written as
\beq
\label{bs}
\psi=\alpha\gamma|00\rangle
+\sqrt{2}\beta\gamma|10\rangle +\frac{\beta\delta}{\sqrt{2}}(|20\rangle-|02\rangle),
\eeq
where $|ij\rangle$ denotes the state
with $i$ particles in mode $a'$ and $j$ particles
in mode $b'$.
Thus, the probability of measuring $|10\rangle$  is
$2|\beta \gamma|^2\le 1/2$, because $|\gamma|=|\alpha|$
from Eq. (\ref{condition}).
Similarly, if the input state is $\alpha|0\rangle-\beta|1\rangle$,
the roles of the $a'^{\dagger}$ and $b'^{\dagger}$ terms are
interchanged and
we obtain the $|01\rangle$ term instead of the $|10\rangle$ term.
In this way, the observers are able to measure
 a projection on a superposed  state
 $\alpha|0\rangle\pm\beta|1\rangle$
 ($P'_A$ and $P'_B$ in Eq. (\ref{P}))
  of a single photon and the vacuum.
We now discuss how to detect the presence of an eavesdropper using
the projective measurement devices described above
in conjunction with 
Bell's inequality.
Choosing four projection operators
\beqa
\label{P}
P_A&\equiv &|1\rangle_A\langle1|_A,~P_B\equiv|1\rangle_B\langle1|_B,\no
P'_A&\equiv&(\alpha|0\rangle_A+ \beta|1\rangle_A)
(\alpha^*\langle 0|_A+ \beta^*\langle 1|_A),\no
P'_B&\equiv&(\alpha|0\rangle_B- \beta |1\rangle_B)
(\alpha^*\langle 0|_B- \beta^* \langle 1|_B),
\eeqa
one can obtain
expectation values of the operators
\beqa
\label{expectation}
\langle \phi | P_A' |\phi \rangle
 &=&\langle \phi| P_B' |\phi\rangle=\frac{1}{2},\no
\langle \phi| P_A' P_B |\phi\rangle &=&
\langle \phi| P_A P_B' |\phi \rangle=\frac{|\beta|^2}{2},\no
\langle \phi| P_A P_B |\phi\rangle &=& 0,~
\langle \phi| P_A' P_B' |\phi \rangle=2|\alpha\beta|^2.
\eeqa
From these expectation values one can define a  quantity
\beqa
\label{S}
S&\equiv&  \langle \phi| P_A' +P_B'-P_A' P_B'
-P_A' P_B-P_A P_B'
+P_A P_B |\phi   \rangle \no
&=&|\alpha|^2(1-2|\beta|^2),
\eeqa
which violates
the following version of Bell's inequality, formulated by Peres 
\beq
\label{bell}
0 \le S \le 1
\eeq
when $|\beta|>1/\sqrt{2}$ and $\alpha\neq0$.
This inequality is obtained when we assume a local hidden variable.
As usual, possible interception, detection and substitution of the
 photons by an eavesdropper
is equivalent to introducing a local hidden variable into the system.
In this case, Alice and Bob obtain not $S$ but
\beqa
\label{SE}
S_E&=&\int \rho(E_A, E_B)dE_A dE_B
 [ p_A(E_A,A')  + p_B(E_B,B')\no
&-&p_A(E_A,A') p_B(E_B,B')
 -p_A(E_A,A') p_B(E_B,B) \no
&-& p_A(E_A,A) p_B(E_B,B') + p_A(E_A,A) p_B(E_B,B)],
\eeqa
where  $\rho(E_A, E_B)$
is the probability that  Eve measures the  projection
 on a state $|E_A\rangle$ at photon A
($P_{|E_A\rangle}$)
and $|E_B\rangle$ at photon B ($P_{|E_B\rangle}$).
This represents
the strategy of the eavesdropper.
$p_A(E_A,A')$ denotes the probability of a count
from Alice's detector
when she tests the projection operator $P'_{A}$
after Eve has tested the projection operator $P_{| E_A \rangle} $
on the photon $A$.
It is expressed by the quantum calculation
\beq
\label{pa}
p_A(E_A,A')=
\langle \phi | P'_A P_{|E_A\rangle} |\phi\rangle.
\eeq
For example, setting $\alpha=1/2$ and $\beta=\sqrt{3}/2$
and considering the special case in which the eavesdropper
measures only photon A, we obtain from Eqs. (\ref{P}) and
(\ref{pa})
\beqa
S_E&=&\int \rho(E_A, E_B)dE_A dE_B
 \left[ 1-p_A(E_A,A') \right]\no
 &=& \int \rho(E_A)dE_A \left [1-|
 \alpha'+ \frac{\sqrt{3}}{2}\beta'|^2 \right] ,
\eeqa
where $|E_A\rangle\equiv\alpha'|0\rangle_A+ \beta'|1\rangle_A$.
With the triangle inequality, this implies $1/4\le S_E \le1$, which
contradicts the quantum prediction of $S=-1/8$
obtained from Eq. (\ref{S}) for the system with no eavesdropper.
In this respect one may say that
 our scheme represents another experimental
method for examining
the single particle nonlocality.

We may now proceed to the discussion of  a key distribution scheme
going as follows.\\
(i) The photon source ($S$) and beam splitters ($BS_0$)
periodically generate the single photon entangled state.
\\
(ii) At a photon arrival time, Alice measures a projection
operator randomly chosen between $P_A$ and $P'_A$.
Similarly, at the same time, Bob  measures $P_B$ or  $P'_B$.
This corresponds to the selection of the analyzer axis in
ordinary two particle quantum cryptography schemes.\\
(iii) After a series of measurements Alice and Bob
announce to each other which projection operator
they chose.
If Alice chose $P_A$ and Bob chose $P_B$
(probability  1/4),
one of them will detect a photon and the other will not.
Then they can share a random raw key $1$
(say, for a photon)  and $0$ (for vacuum).
With a probability of 3/4, either Alice  chooses  $P'_A$ or Bob chooses
$P'_B$. Since their results are not anti-correlated
(see Eq. (\ref{expectation}))
in these cases,
they cannot extract keys. However, these discarded data
together with the anti-correlated data from the previous step
 can be used
to detect eavesdroppers, as shown in the next step. \\
(iv) Detection of eavesdroppers is possible by publicly
comparing a subset of  the  results of Alice and Bob
using Eq. (\ref{S}) and Eq. (\ref{bell}),  as described above.

We now briefly discuss another scheme that adopts
 deterministic projective measurement devices using cavity QED. The
setup of this scheme, shown in Fig. 3,
is similar to that considered by  Davidovich et al.,
Freyberger, Moussa
and Baseia \cite{davidovich,moussa,freyberger},
except that  the single particle entangled state $|\phi\rangle$
is generated not by
an  atom crossing the two cavities, but by the beam
splitter (BS)  and  the single photon source (S) as in Fig. 1.
\begin{figure}[Fig3]
\epsfysize=6cm \epsfbox{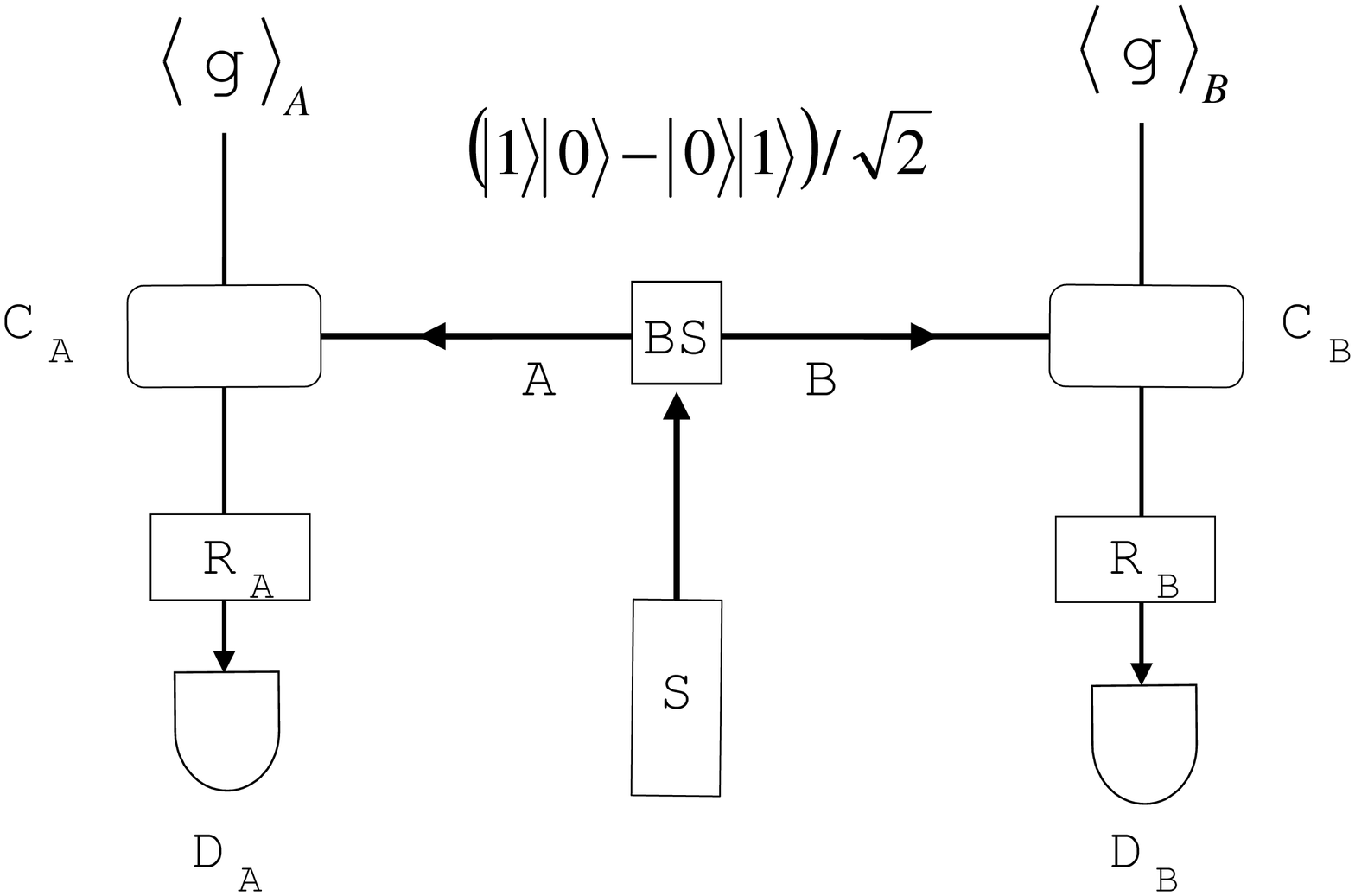}
 \caption[Fig3]{\label{fig3}
Schematic of the apparatus
 for quantum cryptography with deterministic projective measurement
 using cavity QED.
}
\end{figure}
Assuming  that at time $t=0$ two ground state atoms $|g\rangle_A$ and $|g\rangle_B$
are injected into the cavities $C_A$ and $C_B$, respectively,
the total cavities-atom state is then
$|\psi(0)\rangle=|\phi\rangle|g\rangle_A|g\rangle_B$.
The interaction between atoms and photons
in the cavity $C_k ~(k=A,B)$ is described by
the Jaynes-Cummings Hamiltonian
\begin{equation}
H_I^k=\hbar \lambda(\sigma_{+,k}a_k +\sigma_{-,k} a_k^\dagger ),
\end{equation}
where $\lambda$ is a coupling constant
and $\sigma_{+,k},\sigma_{-,k}$ and $a^\dagger_k,a_k$
are the raising and lowering operators for the atom and photon
states, respectively.
In the cavities, these atoms interact with the photons injected into the cavities.
In Refs. \citen{davidovich,moussa,freyberger}, it was shown that
by choosing the interaction time $t$ to be $\lambda t=\pi/2$,
one can replicate the information of the entanglement of
the photon
states $|\phi\rangle$
 to that of the atoms.
 In other words, the state becomes
\beqa
|\psi(t)\rangle&=&exp[-i/\hbar (\Sigma_k H^k_I)t] |\psi(0)\rangle \no
&=&\frac{1}{\sqrt{2}}\left (|e\rangle_A|g\rangle_B
-|g\rangle_A|e\rangle_B \right )|0\rangle_A |0\rangle_B
\label{outputstate},
\eeqa
The projective measurement
on $\alpha|0\rangle+\beta|1\rangle$
 can be performed as follows.
Microwave fields are  appropriately adjusted in the Ramsey
zones ($R_k$) 
such that a superposition  of the ground state
and the excited state of the atom,
$\alpha|g\rangle_k+\beta|e\rangle_k$, with $|\alpha|^2+|\beta|^2=1$,
undergoes a unitary evolution to the  excited state $|e\rangle_k$,
which  registers a click in  the state-selective
 ionization detector $D_k$. 
Except for the measurement devices, the procedure followed in this scheme
 is the same as that with linear optics devices shown in Fig. 1.

Our scheme has the following merits compared to ordinary quantum
cryptography schemes.
First, compared to the ordinary two-particle EPR-based scheme,
it is easier for our scheme to generate vacuum and single particle entanglement
using beam splitters.
Of course, our model entails the detection of a superposition of the vacuum and single photon states, which is rather difficult to implement. 
However, the difficulty involved in detecting the superposed state will also be encountered by eavesdroppers.
Second, compared to non-EPR based schemes such as
the BB84 scheme, it is easier for EPR-based schemes to use quantum repeaters\cite{repeater}
based on quantum teleportation\cite{leekim} to send information
to distant observers.
One shortcoming of our scheme is that,
due to low detection efficiency , Bob may sometimes confuse
a loss of signal with
the vacuum state. In this case, Alice and Bob need to distill a secret key from the
 series of keys using privacy amplification\cite{privacy}.

In summary, we have proposed a new quantum cryptography technique
based on single particle entanglement using linear optics devices
and Bell's inequality
to detect the presence of eavesdroppers.
\\

 We acknowledge the
support of the Korean Ministry of
Education.


\end{document}